# Comment on "An improved upper limit on the neutrino mass from a direct kinematic method by KATRIN"

Alan Chodos
*Department of Physics, University of Texas at Arlington*
*alan.chodos@uta.edu*

**Abstract**: We note that the central value of the KATRIN measurement has negative mass squared, and wonder why the statistical analysis excludes such values *a priori*.

**Introduction, discussion and conclusions**

The new upper limit on the electron-neutrino mass, recently reported by the KATRIN experiment [1], is noteworthy for its increased accuracy, but also for the fact that the measured central value for the mass-squared is negative, in accord with a long tradition that extends back several decades [2,3].

Of course, the central value is barely a standard deviation away from zero, so it is quite correct to quote an upper limit and not a measurement of non-zero mass. However, as stated in the report, the negative mass-squared region is excluded *a priori* as unphysical in performing the statistical analysis, following the practice of most of the authors cited above.

The purpose of this brief remark is to stress that this is an inappropriate restriction. Whether the neutrino does or does not have negative mass-squared is an experimental question. At the very least, the authors should include an alternative analysis in which the possibility of negative neutrino mass-squared is allowed.

Theoretical work on spacelike neutrinos dates back to the mid-1980s [4,5]. More speculative ideas [6] may or may not prove relevant, but regardless, the possibility that the neutrino is a tachyon remains open and should not be discarded even before the analysis begins.